\documentstyle[12pt]{article}
\begin{document}
\baselineskip=13pt
\begin{center} 
{\bf Ising Model on periodic and quasi-periodic chains in presence of magnetic field:some exact results}
\end{center}
\vspace{1.0cm} 
\begin{center}
{\bf Susanta Bhattacharya}$^a$ and {\bf Samir K. Paul}$^b$\\
$^a$ Ramsaday College\\
Amta,Howrah,West Bengal,India\\

$^b$ S. N. Bose National Centre For Basic Sciences\\
Block-JD, Sector-III, Salt Lake\\
Calcutta-700091,  India
\end{center}

\vspace{0.5cm}

\noindent        {\bf Abstract}

   We present a general procedure for calculating the exact partition function of an Ising model on a periodic chain in presence of magnetic field considering both open and closed boundary conditions . Using same procedure on a quasiperiodic (Fibonacci) chain we have established a recurrence relation 
among partition functions of different Fibonacci generations from $n-th$ to $(n+6)-th$ . In the large $N$ limit we find $(2\tau +1){F_{n+1}}={F_{n-2}}$ ; where $\tau$ is the golden mean and $F_n$ stands for free energy/spin for the $n-th$ generation . We have also studied chemical potential in both cases.
\vspace{0.2cm}

\noindent PACS No: 05.50.+q
 
\vspace{0.5cm}

\noindent  a)  susanta1005@rediffmail.com

\noindent  b)  smr@boson.bose.res.in

\vspace{0.5cm}

\noindent        {\bf {I Introduction}}

\vspace{0.2cm}

In this paper we have calculated the exact partition function of an Ising model on a periodic chain in presence of magnetic field with open boundary conditions by splitting the transfer matrix into two particular non-commuting matrices.    
The result is quite nontrivial in contrast to the expression for the partition  function of the closed one[1,2]. Since there has been an enormous amount of work on Ising model it is difficult to do justice to all references, however rigorous results and developments can be found in [1,3] and in the review article [4] .We have studied chemical potentials for periodic chain in presence of  magnetic field . Studying Ising model on a Fibonacci chain in presence of magnetic field is interesting since the discovery of quasicrystals [5]. Scaling forms of thermodynamic functions for such system have been studied using renormalisation group technique [6] through one step decimation . The ground state and thermodynamic properties of such a system have been studied using renormalization group technique [6,7] . Our formulation      
 helps us to express the partition function of Ising model on Fibonacci chain in presence of magnetic field as a sum  of partition  functions of usual Ising open chains with coefficients containing Fibonacci symmetry.We have also studied some symmetry properties of the Fibonacci chain.Using a special symmetry property ("Mirror Symmetry") and the usual trace map relation [8,9] ( trace maps and invariants relating to two-letter substitution lattices have been studied in [10] and references therein ) we have established a 
recurrence relation among the partition functions of different Fibonacci
generations.This includes all the partition functions starting from $n-th$ up to
$(n+6)-th$ generations.We observe that mirror symmetry is a characteristic property of each Fibonacci generation with $n-th$ and $(n+6)-th$ generations having  
same topology [11]. Assuming that the free energy/spin for both open and closed chains for a particular generation are equal in the large $N$ limit we have obtained a scaling property among the free energy/spin of two consecutive even genarations, using the above recurrence relation . In this particular case we have studied chemical potential in absence of magnetic field.  

\vspace{1.0cm}

\noindent {\bf {II Exact partition function for open Ising chain with magnetic field}}

\vspace{0.2cm}

  The one dimensional Ising model consists of a chain of N spins ${S_i}={\pm 1}$

 ;$i =1,2,.....,N$ with nearest neighbour interactions ${\epsilon_{i,i+1}}$. The

 Hamiltonian is given by:

\begin{equation}
{\mathcal{H}} = - {\sum_{i=1}^{N-1}}{\epsilon_{i,i+1}} {S_i} {S_{i+1}} - H{\sum_{i=1}^{N}}{S_i}
\end{equation}

For a uniform lattice ${\epsilon_{i,i+1}} = {\epsilon}$,the partition function   is given by: 
 
\begin{equation}
{Z_N^o}(T,H) = {\sum_{{S_1},{S_2},....,{S_N}= - 1}^{+1}} f({S_1},{S_2})
f({S_2},{S_3})......f({S_{N-1}},{S_N}) {f_0}({S_N},{S_1})
\end{equation}

with
$f({S_i},{S_{i+1}}) =exp[{\beta}{\epsilon}{S_i}{S_{i+1}} + {\frac{1}{2}}
\noindent       {\beta}H({S_i}+{S_{i+1}})]$; 
${f_0}({S_N},{S_1}) = {{[f({S_N},{S_1})]}_{\epsilon =0}}$.                      \noindent   Here the superscript $o$ stands for the chain with open
boundary condition.Therefore the partition function (2) can be written in terms
of transfer matrix as: 

\begin{equation}
{Z_N^o}(T,H) = Tr {P^{N-1}}{P_0}
\end{equation}

where
\begin{equation}
P = {\sqrt{r}}(1+{\frac{\lambda}{r}}){{\sigma}_1} = {\sqrt{r}}{{\sigma}_1}(1+
{\frac{{\lambda}^T}{r}})
\end{equation}

\begin{equation}
{P_0}={[P]_{\epsilon =0}}=(1+{\lambda}){\sigma_1}={\sigma_1}(1+{\lambda^T})
\end{equation}
 
with 
$r = exp(-2{\beta}{\epsilon})$ , 
$\lambda = \left(\matrix{0 & e^{\beta H}\cr e^{-\beta H} & 0}\right),$
${\lambda^T} = \left(\matrix{0 & e^{-\beta H}\cr e^{\beta H} & 0}\right)$
and ${\sigma_1} = \left(\matrix{0 & 1 \cr 1 & 0}\right).$ 

\vspace{0.2cm}

 The general formula of the partition function for even and odd number

of spins (i.e.,odd and even number of bonds) can be derived by using equations 

(3) and (4) as :

\begin{equation}
{Z_{2N}^o}(T,H) = {r^{N-{\frac{1}{2}}}}Tr(1+{x_1})(1+{x_2})....(1+{x_{2N-1}})
(1+{\lambda^T})
\end{equation}

 and 

\begin{equation}
{Z_{2N+1}^o}(T,H) = {r^N}Tr(1+{x_1})(1+{x_2})......(1+{x_{2N}})(1+\lambda )
{\sigma_1}
\end{equation}
 
where 

\begin{equation}
{x_{2i+1}} = {\frac{\lambda}{r}},     {x_{2i}} = {\frac{\lambda^T}{r}}; i=integer     
\end{equation}

The above equations show that $\lambda , {\lambda^T}$ are the signatures for
the transfer matrices corresponding to bonds in odd and even positions.In the
case of a chain with closed boundary condition the last factor in eqn.(2) is
$f({S_N},{S_1})$ and consequently the partition function takes the form: 

\begin{equation}
{Z_{2N}^c}(T,H) = {r^N} Tr(1+{x_1})(1+{x_2})......(1+{x_{2N}})
\end{equation} 

\begin{equation}
{Z_{2N+1}^c} = {r^{N+{\frac{1}{2}}}}Tr(1+{x_1})(1+{x_2}).....(1+{x_{2N+1}})
{\sigma_1}
\end{equation} 

Here the superscript $c$  indicates closed chain.One can show by elementary calculation that eqns.(9) and (10) reduce to the well known form[12] 

\begin{equation}
{Z_N^c}(T,H) = {\lambda_{+}^N} + {\lambda_{-}^N}
\end{equation} 

where 

\begin{equation}
{\lambda_{\pm}} = {r^{-\frac{1}{2}}}[{\cosh{(\beta H)}} \pm {\sqrt{({{\sinh}^2}
{(\beta H)} + {r^2})}}]
\end{equation} 

are the eigenvalues of the transfer matrix $P$.The expression for the partition  function in the case of an open chain with even number of spins can be derived from eqn.(6) as follows: 

\begin{eqnarray} 
{Z_{2N}^o}(T,H) 
 = {r^{N-{\frac{1}{2}}}}Tr(1+{x_1})(1+{x_2}).....(1+{x_{2N-1}})(1+\lambda^T)\nonumber\\
 = {\sqrt{r}}{Z_{2N}^c}(T,H) + {r^{N-{\frac{1}{2}}}}(1-r)Tr(1+{x_1})(1+{x_2})
.......(1+{x_{2N-1}})\nonumber\\
 = {\sqrt{r}}{Z_{2N}^c}(T,H) + {\sqrt{r}}(1-r){Z_{2(N-1)}^c}(T,H) + 
{r^{N-{\frac{1}{2}}}}(1-r)\nonumber\\
\times Tr (1+{x_1})(1+{x_2}).......(1+{x_{2N-2}}){x_{2N-1}}
\end{eqnarray}

The last term in the above expression can be written in terms of the eigenvalues of the transfer matrix $P$ viz.${\lambda_\pm}$. By following the method of induction :

\begin{eqnarray} 
{r^{N-{\frac{1}{2}}}}(1-r)Tr(1+{x_1})(1+{x_2})......(1+{x_{2N-2}}){x_{2N-1}}
\nonumber\\
= (1-r){r^{N-{\frac{1}{2}}}}{\frac{4}{r^2}}{\cosh^2}{(\beta H)}
{\sum_{i=0}^{N-2}}{\bigg( {\frac{{\lambda_+}^2}{r}} \bigg)}^{N-2-i}   
 {\bigg( {\frac{{\lambda_-}^2}{r}} \bigg)}^{i}\nonumber\\                                   
=4(1-r){r^{-\frac{1}{2}}}{\cosh^2}(\beta H)\frac{{{\lambda_+}^{2(N-1)}}-{{\lambda_-}^{2(N-1)}}}{{{\lambda_+}^2}-
{{\lambda_-}^2}}
\end{eqnarray} 

So eqn.(13) becomes : 

\begin{eqnarray}
{Z_{2N}^o}(T,H)& = &{\sqrt{r}}{Z_{2N}^c}(T,H)+{\sqrt{r}}(1-r){Z^c_{2(N-1)}}(T,H)\nonumber\\                                                                     &  &+ 4(1-r){r^{-\frac{1}{2}}}{\cosh^2}
{(\beta H)}
\times {\frac{{{\lambda_+}^{2(N-1)}}-{{\lambda_-}^{2(N-1)}}}{{{\lambda_+}^2}
- {{\lambda_-}^2}}}
\end{eqnarray}
 
Similarly the expression (7) for the open chain partition function with odd
number of spins takes the form:

\begin{eqnarray}
{Z_{2N+1}^o}(T,H) = {\sqrt{r}}{Z_{2N+1}^c}+4(1-r){\cosh{(\beta H)}}\nonumber\\
\times \frac{{{\lambda_+}^{2N}}-{{\lambda_-}^{2N}}}{{{\lambda_+}^2}-{{\lambda_+}^2}}
\end{eqnarray} 

It can be shown easily that the free energy per spin calculated from equations (11) , (15) and (16) are all same.
Thus the forms of the thermodynamic functions are  same for both open and closed  chains . However as the expressions of the partition function are different for closed and open chain it will be highly instructive to study the chemical potential of such systems.

\vspace{0.2cm} 

\noindent {\bf IIa Chemical Potential}

\vspace{0.1cm}

The chemical potential for an open chain with $N$ number of spins is given by:
\begin{equation}
{\mu^o_N}={\frac {d~log{Z^o_N}}{d~N}}= log{\frac{Z^o_{N+1}}{Z^o_N}} 
\end{equation} 
The expression for the chemical potential for a system with even number of spin ' say $2N$ ; is obtained by adding a spin to the spin to the system and substituting from equations (15) and (16) in the above relation (17) in the large $N$ limit:
\begin{equation}
{e^{\mu^o_{even}}\\
 = {\nu^o_{even}}\\     
 = {\frac{Z^o_{2N+1}}{Z^o_{2N}}} \\
 = {({\nu_c})^2}\times {\frac{{\sqrt r}({\nu_c})+4(1-r)cosh({\beta H})\times {\frac{1}{{{\lambda_+}^2}-{{\lambda_-}^2}}}}{{\sqrt r}{({\nu_c})^2}+{\sqrt r}(1-r)+4{r^{\frac{-1}{2}}}(1-r){{cosh}^2}{\beta H}{\frac{1}{{{\lambda_+}^2}-{{\lambda_-}^2}}}}}}
\end{equation}
where ${\nu_c}={e^{\mu_c}}={\frac{Z^c_{N+1}}{Z^c_N}}={\lambda_+}$ , $\mu_c$ being the chemical potential for the closed chain.For $H=0$, ${\lambda_+}=2cosh~{\beta \epsilon}$ and consequently expression (18) becomes
\begin{equation}
{\nu^o_{even}}=2cosh~{\beta \epsilon}
\end{equation}
Similarly for a system with odd number of spins we can write
\begin{equation}
{e^{\mu^o_{odd}}}={\nu^o_{odd}}={\frac {Z^o_{2N}}{Z^o_{2N-1}}}={({\nu_c})^2}\times {\frac{1}{\nu^o_{even}}}
\end{equation}
Therefore, ${\nu^o_{even}}\times {\nu^o_{odd}}={({\nu_c})^2}$ which implies 
\begin{equation}
{\mu^o_{even}} + {\mu^o_{odd}} = 2{\mu_c}
\end{equation}
 From the above expressions we conclude that the chemical potentials $\mu_c$,$\mu^o_{even}$,$\mu^o_{odd}$ are all different. However,for
$H=0$
\begin{equation}
{\mu_c}={\mu^o_{even}}={\mu^o_{odd}}=log~(2~cosh~{\beta \epsilon})
\end{equation}
which shows that the chemical energies for three different conditions of the chain mentioned above are
degenerate and it is removed when magnetic field is applied [equation (21)].

\vspace{0.5cm}

\noindent {\bf {III  Ising model on Fibonacci chain with magnetic field}}

\vspace{0.2cm}

  A Fibonacci chain can be inflated by two bonds $L(large)$ and $S(small)$ by the
inflation rule $L \longrightarrow LS,S \longrightarrow L$.The chain can be
represented by the sequence: 

\begin{equation}
L \longrightarrow LS \longrightarrow LSL \longrightarrow LSLLS \longrightarrow 
 LSLLSLSL \longrightarrow .....
\end{equation}

In this case the interaction strengths in the Hamiltonian (1) ${\epsilon_{i,i+1}}=\epsilon$ for long bonds and ${\epsilon_{i,i+1}}={\bar \epsilon}$ for the short ones where the bonds are arranged according to the Fibonacci sequence (23).The corresponding partition function of the $nth$ generation Fibonacci chain having 
$N$ spins with $N-1$ bonds is given by:

\begin{equation}
{Z_N^o}(F)=Tr P{\bar P}PP{\bar P}.....{P_0}
\end{equation}

where for long bonds the transfer matrix $P$ is given by eqn.(4) and for short 
bonds the transfer matrix  ${\bar P}$ is given by eqn.(4) with $r$ replaced by
${\bar r}= {r|}_{\epsilon = {\bar \epsilon}}$.Henceforth ${Z_N^o}(F)$ and ${Z_N^o}(I)$ will represent partition functions for  Ising models on an open Fibonacci
chain and on an open regular lattice respectively.The expressions for the partition functions with odd and even number of bonds take the same forms as shown in eqns.(6) and (7) with $x_i$'s given in eqn.(8) for long bonds whereas for short bonds
we replace $r$ by ${\bar r}$ in eqn.(8).The explicit expressions for the partition functions for open and closed chains are: 

\begin{equation}
{Z_{2N}^o}(F)={r^{\frac{N_L}{2}}}{{\bar r}^{\frac{N_S}{2}}}Tr (1+{x_1})(1+{x_2})......(1+{x_{2N-1}})({1+\lambda^T})
\end{equation}
    
\begin{equation}
{Z_{2N+1}^o}(F)={r^{\frac{N_L}{2}}}{{\bar r}^{\frac{N_S}{2}}}Tr (1+{x_1})(1+{x_2})......(1+x_{2N})({1+\lambda}){\sigma_1}
\end{equation}

Note that the subscript $2N$ in the left hand side of eq.(25) stands for number of spins so that number of bonds is $2N-1$ which is odd.In
eq.(26) number of spins is $2N+1$ and number of bonds is $2N$ which is even.

Similarly for closed chain eqs. (24),(25) and (26) take the following forms:

\begin{equation}
{Z_N^c}(F)= Tr P{\bar P}PP{\bar P}.......P
\end{equation}

\begin{equation}
{Z_{2N}^c}(F)={r^{\frac{{N_L}+1}{2}}}{{\bar r}^{\frac{N_S}{2}}}Tr (1+{x_1})(1+{x_2})......(1+{x_{2N-1}})(1+{x_{2N}})
\end{equation}

\begin{equation}
{Z_{2N+1}^c}(F)={r^{\frac{{N_L}+1}{2}}}{{\bar r}^{\frac{N_S}{2}}}Tr(1+{x_1})(1+{x_2}).......(1+{x_{2N}})(1+{x_{2N+1}}){\sigma_1}
\end{equation}

where $N_L$,$N_S$ are number of long and short bonds in a particular sequence.

\vspace{0.1cm} 
 
Now ${\bar P}$ is related to P through the following equation:

\begin{equation}
{\bar P}={\sqrt{\bar r}}(1-{\frac{r}{\bar r}}){\sigma_1}+{\sqrt{{\frac{r}{\bar r}}}}   P
\end{equation}

Using eqn.(30) in eqns.(25) and (26) the Fibonacci partition function for any generation can be written in terms of open Ising partition functions as follows:

\begin{equation}
{Z_{2N}^o}(F)={h_0}({\epsilon,{\bar \epsilon}})+{\sum_{i=1}^N}{h_{2i}}({\epsilon, {\bar \epsilon}}){Z_{2i}^o}(I)
\end{equation}

\begin{equation}
{Z_{2N-1}^o}(F)={l_0}({\epsilon,{\bar \epsilon}})+{\sum_{i=1}^N}{l_{2i-1}}({\epsilon,{\bar \epsilon}}){Z_{2i-1}^o}(I)
\end{equation}

where ${Z_{2i}^o}(I)$ and ${Z_{2i-1}^o}(I)$ are given by eqns.(15) and (16) respectively.We observe that the quasiperiodic nature of the Fibonacci chain is encoded in the functions $h(\epsilon,{\bar \epsilon})$ and $l(\epsilon,{\bar \epsilon})$.Though for small generations these functions can be derived exactly still we could not find out their general forms. 

\vspace{0.2cm}

\noindent  {\bf  {IIIa Recurrence relation among partition functions}}

\vspace{0.1cm}

To circumvent the above difficulty we study the recurrence  relations among the partition functions of different Fibonacci generations.A servey of different Fibonacci generations depicted by eqn.(23) shows a symmetric pattern in terms of the number of bonds ,viz.,

\begin{eqnarray}
{P_1} : P (odd)\nonumber\\
{P_2} : P{\bar P}(even)\nonumber\\
{P_3} : P{\bar P}P(odd)\nonumber\\
{P_4} : P{\bar P}PP{\bar P}(odd)\nonumber\\
{P_5} : P{\bar P}PP{\bar P}P{\bar P}P(even)\nonumber\\
{P_6} : P{\bar P}PP{\bar P}P{\bar P}PP{\bar P}PP{\bar P}(odd)
\end{eqnarray}
and so on.\\

 The trace map relation [8,9] was introduced to study the spectrum of 1D Schrodinger equation in a discontinious quasiperiodic potential.The use of trce map in different kind of substitution lattices have been studied in the review article   [4] with many relevant references therein.

\vspace{0.2cm}

    Let $P_{n-2}$ be the $(n-2)th$  Fibonacci generation with even number of bonds.This automatically ensures that the previous as well as the next two consecutive generations will have odd number of bonds . The recurrence  relation for   Fibonacci generations is given by:

\begin{equation}
{P_n}={P_{n-1}}{P_{n-2}}
\end{equation}

Now adding a term ${D_{n-2}}{P_{n-3}}$ in the above equation gives

\begin{equation}
{P_n} + {D_{n-2}}{P_{n-3}} = {P_{n-1}}{P_{n-2}} + {D_{n-2}}{P_{n-3}}
\end{equation}

where ${D_{n-2}}=Det({P_{n-2}})$.The following operations are applied            sequentially on eqn.(35): 

\vspace{0.1cm}

     substitute ${{P_{n-2}}^{-1}}{P_{n-1}}$ in place of ${P_{n-3}}$ on
the right hand side and finally use Cayley-Hamilton theorem to get the usual trace map relation [8] on the Fibonacci lattice:

\begin{equation}
Tr {P_n} = Tr {P_{n-1}} Tr {P_{n-2}} - {D_{n-2}}Tr {P_{n-3}}
\end{equation}
 
 The above equation will be necessary for calculating recurrence relation among
different Fibonacci generations.
For this purpose  we must      understand symmetry properties of Fibonacci chain.Inspecting different generations of the Fibonacci chain it reveals that if the  total number of bonds $N$ of a particular generation is odd then there is a 
 mirror reflection symmetry  arround the ${\bigg( {\frac{N-1}{2}} \bigg)
}th$ bond ;except the last two bonds.If the special bond arround which mirror symmetry
 occurs is a short(long) one the Fibonacci generation will have equal number of  odd and even short(long) bonds . However if the total number of bonds $N$ is
even,the mirror reflection symmetry is arround a cluster of two successive long  bonds at the ${\bigg( {\frac{N}{2}} \bigg)}th$ and ${\bigg( {\frac{N-2}{2}} \bigg)}th$ positions of the chain . 
So "Mirror reflection symmetry"  is a characteristic property of a Fibonacci
chain.

\vspace{0.1cm}

 The $nth$ and $(n\pm 3)th$ generations have the  mirror reflection symmetry property arround the same kind of bond with last two bonds interchanged , while the $(n\pm 6)th$ generations are topologically same as the $nth$ one.

\vspace{0.2cm}

   Using recurrence relation (34) we can write

\begin{equation}
{D_{n-2}}{P_{n-3}}={D_{n-2}}{{P_{n-2}}^{-1}}{P_{n-1}}
\end{equation}

Using Cayley-Hamilton theorem on the right hand side of eq.(37) we get:

\begin{equation}
{D_{n-2}}{P_{n-3}}=(Tr{P_{n-2}}){P_{n-1}}-{P_{n-2}}{P_{n-1}}
\end{equation}

Multiplying eq.(38) by P from the right and taking trace we obtain:

\begin{equation}
{D_{n-2}}{Z^c_{n-3}}=(Tr{P_{n-2}}){Z^c_{n-1}}-Tr({P_{n-2}}{P_{n-1}}P)
\end{equation}

In a  similar fashion we obtain :

\begin{equation}
{D_{n-2}}{Z^o_{n-3}}=(Tr{P_{n-2}}){Z^o_{n-1}}-Tr({P_{n-2}}{P_{n-1}}{P_0})
\end{equation}

The avove relations cleary show that one cannot obtain a recurrence relation
among the partition functions through the trace map relation alone.To circumvent this difficulty we take recourse to mirror symmetry property. 
The expression ${P_{n-2}}{P_{n-1}}$ in eqns.(39) and (40) is similar to 
${P_n}={P_{n-1}}{P_{n-2}}$ with last two bonds interchanged,i.e.,both of them
have the same mirror symmetric part $\Omega_n$.Therefore eqns.(39) and (40) can
 be written as: 

\begin{equation}
{D_{n-2}}{Z^c_{n-3}}={Z^c_{n-1}}(Tr{P_{n-2}})-Tr({\Omega_n}P{\bar P}P)
\end{equation}

\begin{equation}
{D_{n-2}}{Z^o_{n-3}}={Z^o_{n-1}}(Tr{P_{n-2}})-Tr({\Omega_n}P{\bar P}{P_0})
\end{equation}

The transfer matrix has the property that $P={P^T}$ and ${\bar P}={{\bar P}^T}$
.If such transfer matrices are arranged in a mirror symmetric fashion then the resulting matrix $({\Omega_n})$ will have the following properties:

\vspace{0.2cm}

i) Off diagonal elements are same i.e.,${({\omega_n})_{12}}={({\omega_n})_{21}}$

\vspace{0.1cm}

ii)Diagonal elements are not same but satisfy the condition:

\vspace{0.1cm}

${({\omega_n})_{11}}(p,q)={({\omega_n})_{22}}(q,p)$;                            where $p={e^{{\beta}H}}$,$q={e^{-{\beta}H}}$. 

\vspace{0.1cm}

Thus the matrix ${\Omega_n}$ in
eqns. (41) and (42) is of the form: 

\vspace{0.2cm}

${\Omega_n}=\left(\matrix {{({\omega_n})_{11}} & {({\omega_n})_{12}}\cr 
{({\omega_n})_{21}} & {({\omega_n})_{22}}}\right)$ 

\vspace{0.2cm}

Eqns.(41) and (42) can be written explictly in the following way:

\begin{eqnarray}
{D_{n-2}}{Z^c_{n-3}}& = & {Z^c_{n-1}}(Tr{P_{n-2}})
-r{\sqrt{\bar r}}[{({\omega_n})_{11}}(y+{\frac{pu}{r}})+{({\omega_n})_{12}}(u+v+{\frac{px+qy}{r}})\nonumber\\
+{({\omega_n})_{22}}(x+{\frac{qv}{r}})]
\end{eqnarray}

\begin{eqnarray}
{D_{n-2}}{Z^o_{n-3}}& = & {Z^o_{n-1}}(Tr{P_{n-2}})
-{\sqrt{r{\bar r}}}[{({\omega_n})_{11}}(y+pu)+{({\omega_n})_{12}}(u+v+px+qy)\nonumber\\
+{({\omega_n})_{22}}(x+qv)]
\end{eqnarray}

where we have used

\vspace{0.1cm}

\hspace{0.5cm}

$P{\bar P}={\sqrt{r{\bar r}}}\left(\matrix {u & y\cr x & v}\right)$

\vspace{0.1cm}

with 

\vspace{0.2cm} 
 
$x={\frac{p}{\bar r}}+{\frac{q}{r}}$,
$y={\frac{p}{r}}+{\frac{q}{\bar r}}$,$u=1+{\frac{p^2}{r{\bar r}}}$ and
$v=1+{\frac{q^2}{r{\bar r}}}$. 

\vspace{0.2cm}

Eliminating $Tr{P_{n-2}}$ from eqns.(43) and (44) we have:

\begin{eqnarray}
{\frac{y{V_{n-1}}+pu{{V^\prime}_{n-1}}}{(u+v){V_{n-1}}+(px+qy){{V^\prime}_{n-1}}}} {({\omega_n})_{11}}
+{\frac{x{V_{n-1}}+qv{{V^\prime}_{n-1}}}{(u+v){V_{n-1}}+(px+qy){{V^\prime}_{n-1}}}} {({\omega_n})_{22}}
+{({\omega_n})_{12}}\nonumber\\
={\frac{D_{n-2}}{r{\sqrt{\bar r}}}}\times {\frac{{Z_{n-1}^o}{Z_{n-3}^c}-{Z_{n-1}^c}{Z_{n-3}^o}}{(u+v){V_{n-1}}+(px+qy){{V^\prime}_{n-1}}}}
\end{eqnarray}

where 

\vspace{0.2cm}

${V_{n-1}}={\frac{Z_{n-1}^c}{\sqrt{r}}}-{Z_{n-1}^o}$ and
${{V^\prime}_{n-1}}={\frac{Z_{n-1}^c}{\sqrt{r}}}-{\frac{Z_{n-1}^o}{r}}$.

\vspace{0.2cm}

To solve for ${({\omega_n})_{11}}$,${({\omega_n})_{12}}$ and ${({\omega_n})_{22}}$ we need another two equations.These equations are obtained from the usual formulae: 

\vspace{0.2cm}

${Z_n^c}=Tr({\Omega_n}{\bar P}PP)$,   ${Z_n^o}=Tr({\Omega_n}{\bar P}P{P_0})$. 

\vspace{0.2cm}

The explicit forms of these two relations are:

\begin{eqnarray}
{\frac{x+{\frac{pu}{r}}}{u+v+{\frac{qx+py}{r}}}} {({\omega_n})_{11}}
+{\frac{y+{\frac{qv}{r}}}{u+v+{\frac{qx+py}{r}}}} {({\omega_n})_{22}}
+{({\omega_n})_{12}}\nonumber\\
 = {\frac{Z_n^c}{r{\sqrt{\bar r}}}} {\frac{1}{u+v+{\frac{qx+py}{r}}}}
\end{eqnarray}

\begin{eqnarray}
{\frac{x+pu}{u+v+qx+py}}  {({\omega_n})_{11}}
+{\frac{y+qv}{u+v+qx+py}} {({\omega_n})_{22}}+{({\omega_n})_{12}}\nonumber\\
 = {\frac{Z_n^o}{{\sqrt{r{\bar r}}}}} {\frac{1}{u+v+qx+py}}
\end{eqnarray}

By elementary calculation one obtains:

\begin{eqnarray}
{({\omega_n})_{11}}(x,y)=
{\frac{{\Gamma_{n-1}}-{{{\Gamma}^\prime}_{n-1}}}{{\Gamma_{n-1}}{{{\Lambda}^\prime}_{n-1}}-{{{\Gamma}^\prime}_{n-1}}{\Lambda_{n-1}}}}\times {\frac{D_{n-2}}{r\sqrt{\bar r}}}\times {\Delta_{n-1}}\nonumber\\
-{\frac{1}{\sqrt{r{\bar r}}}}{\frac{1}{{\Gamma_{n-1}}{{{\Lambda}^\prime}_{n-1}}-{{{\Gamma}^\prime}_{n-1}}{\Lambda_{n-1}}}}\times ({\Gamma_{n-1}}{Z_n^o}{K^\prime}-{{{\Gamma}^\prime}_{n-1}}{\frac{Z_n^c}{\sqrt{r}}}K)
\end{eqnarray}

\begin{eqnarray}
{({\omega_n})_{22}}(x,y)=-{\frac{{\Lambda_{n-1}}-{{{\Lambda}^\prime}_{n-1}}}{{\Gamma_{n-1}}{{{\Lambda}^\prime}_{n-1}}-{{{\Gamma}^\prime}_{n-1}}{\Lambda_{n-1}}}}\times {\frac{D_{n-2}}{r\sqrt{\bar r}}}\times {\Delta_{n-1}}\nonumber\\
+{\frac{1}{\sqrt{r{\bar r}}}}{\frac{1}{{\Gamma_{n-1}}{{{\Lambda}^\prime}_{n-1}}-{{{\Gamma}^\prime}_{n-1}}{\Lambda_{n-1}}}}\times ({\Lambda_{n-1}}{Z_n^o}{K^\prime}-{{{\Lambda}^\prime}_{n-1}}{\frac{Z_n^c}{\sqrt{r}}}K)
\end{eqnarray}

and 

\begin{equation}
{({\omega_n})_{12}}(x,y)={\frac{1}{(qx+py)(1-{\frac{1}{r}})}}\bigg[ -{\frac{1}{\sqrt{r{\bar r}}}}\times {V_n}-{(1-{\frac{1}{r}})}\bigg( pu{({\omega_n})_{11}}+qv{({\omega_n})_{22}} \bigg) \bigg]
\end{equation}

where

\begin{equation}
{\Lambda_{n-1}}(x,y)={\frac{y{V_{n-1}}+pu{{V^\prime}_{n-1}}}{(u+v){V_{n-1}}+(px+qy){{V^\prime}_{n-1}}}}-{\frac{x+{\frac{pu}{r}}}{u+v+{\frac{qx+py}{r}}}} 
\end{equation}

\begin{equation}
{{{\Lambda}^\prime}_{n-1}}(x,y)={\frac{y{V_{n-1}}+pu{{V^\prime}_{n-1}}}{(u+v){V_{n-1}}+(px+qy){{V^\prime}_{n-1}}}}-{\frac{x+pu}{u+v+qx+py}}
\end{equation}

\begin{equation}
{\Gamma_{n-1}}(x,y)=\frac{x{V_{n-1}}+qv{{V^\prime}_{n-1}}}{(u+v){V_{n-1}}+(px+qy){{V^\prime}_{n-1}}}-\frac{y+{\frac{qv}{r}}}{u+v+{\frac{qx+py}{r}}} 
\end{equation}

\begin{equation}
{{{\Gamma}^\prime}_{n-1}}(x,y)= \frac{x{V_{n-1}}+qv{{V^\prime}_{n-1}}}{(u+v){V_{n-1}}+(px+qy){{V^\prime}_{n-1}}}-\frac{y+qv}{u+v+qx+py} 
\end{equation}

\begin{equation}
{\Delta_{n-1}}(x,y)={\frac{{Z_{n-1}^o}{Z_{n-3}^c}-{Z_{n-1}^c}{Z_{n-3}^o}}{(u+v){V_{n-1}}+(px+qy){{V^\prime}_{n-1}}}} 
\end{equation}

\begin{equation}
K(x,y)={\frac{1}{u+v+{\frac{qx+py}{r}}}}
\end{equation}

\begin{equation}
{K^\prime}(x,y)={\frac{1}{u+v+qx+py}} 
\end{equation}

Eliminating $D_{n-2}$ from eqns.(43) and (44) we obtain:

\begin{eqnarray}
Tr{P_{n-2}}={\frac{r{\sqrt {\bar r}}\times {V_{n-3}}}{{Z^o_{n-1}}{Z^c_{n-3}}-{Z^c_{n-1}}{Z^o_{n-3}}}} \bigg[ \bigg( {\alpha_{xy}}+{\beta_{xy}}{\frac{{V^\prime}_{n-3}}{V_{n-3}}}\bigg){({\omega_n})_{11}}(x,y)\nonumber\\
+\bigg( {{\alpha^\prime}_{xy}}+{{\beta^\prime}_{xy}} {\frac{{V^\prime}_{n-3}}{V_{n-3}}} \bigg){({\omega_n})_{22}}(x,y)\nonumber\\
-{\frac{1}{{\sqrt{r\bar r}}(1-{\frac{1}{r}})(qx+py)}}\bigg( (u+v)+(px+qy){\frac{{V^\prime}_{n-3}}{V_{n-3}}} \bigg) {V_n} \bigg]
\end{eqnarray}

where 

\vspace{0.2cm}

${\alpha_{xy}}=y-{\frac{u+v}{py+qx}}\times pu$, \hspace{0.5cm}          ${\beta_{xy}}={\frac{(x-y)(q-p)}{py+qx}}\times pu$

\vspace{0.2cm}

${{\alpha^\prime}_{xy}}=x-{\frac{u+v}{py+qx}}\times qv$,  \hspace{0.5cm} 
${{\beta^\prime}_{xy}}={\frac{(x-y)(q-p)}{py+qx}}\times qv$  

\vspace{0.2cm}
 
In general $n$th and $(n\pm 2)$th generations have same arrangement of the last two bonds appart from their respective mirror symmetric parts.That is why $Tr{P_n}$ and $Tr{P_{n\pm 2}}$ will have similar expressions.Since we have assumed
${P_n}={\Omega_n}{\bar P}P$ it follows from the expression (33) that ${P_{n-3}}={\Omega_{n-3}}P{\bar P}$.Therefore proceeding in a similar way as above we get:
 
\begin{eqnarray}
Tr{P_{n-3}}={\frac{r{\sqrt{\bar r}}\times {V_{n-4}}}{{Z^o_{n-2}}{Z^c_{n-4}}-{Z^c_{n-2}}{Z^o_{n-4}}}}\bigg[ \bigg( {\alpha_{yx}}+{\beta_{yx}}{\frac{{V^\prime}_{n-4}}{V_{n-4}}}\bigg){({{\omega}_{n-1}})_{11}}(y,x)\nonumber\\
+\bigg( {{\alpha^\prime}_{yx}}+{{\beta^\prime}_{yx}}{\frac{{V^\prime}_{n-4}}{V_{n-4}}} \bigg) {({{\omega}_{n-1}})_{11}}(y,x)\nonumber\\
-{\frac{1}{{\sqrt{r\bar r}}(1-{\frac{1}{r}})(px+qy)}}\bigg( (u+v)+(qx+py){\frac{{V^\prime}_{n-4}}{V_{n-4}}} \bigg) {V_{n-1}} \bigg]
\end{eqnarray}

Using eqns. (58) and (59) and similar expressions for $Tr{P_n}$ , $Tr{P_{n-1}}$ in the trace map relation (36) we obtain the following recurrence  relation among partition functions of different Fibonacci generations as:

\begin{eqnarray}
{\frac{V_{n-1}}{{Z^o_{n+1}}{Z^c_{n-1}}-{Z^c_{n+1}}{Z^o_{n-1}}}} \bigg[ \bigg( {\alpha_{xy}}+{\beta_{xy}} {\frac{{V^\prime}_{n-1}}{V_{n-1}}} \bigg) {({\omega_{n+2}})_{11}}(x,y)+\bigg( {{\alpha^\prime}_{xy}}+{{\beta^\prime}_{xy}}{\frac{{V^\prime}_{n-1}}{V_{n-1}}} \bigg) {({\omega_{n+2}})_{22}}(x,y)\nonumber\\
-{\frac{1}{{\sqrt{r\bar r}}(1-{\frac{1}{r}})(qx+py)}}\bigg( u+v+(px+qy){\frac{{V^\prime}_{n-1}}{V_{n-1}}} \bigg){V_{n+2}} \bigg] ={\frac{r{\sqrt{\bar r}}\times {V_{n-2}}}{{Z^o_n}{Z^c_{n-2}}-{Z^c_n}{Z^o_{n-2}}}}\nonumber\\
\times {\frac{V_{n-3}}{{Z^o_{n-1}}{Z^c_{n-3}}-{Z^c_{n-1}}{Z^o_{n-3}}}} \bigg[ \bigg( {\alpha_{yx}}+{\beta_{yx}}{\frac{{V^\prime}_{n-2}}{V_{n-2}}} \bigg){({\omega_{n+1}})_{11}}(y,x)+ \bigg( {{\alpha^\prime}_{yx}}+{{\beta^\prime}_{yx}} {\frac{{V^\prime}_{n-2}}{V_{n-2}}} \bigg) {({\omega_{n+1}})_{22}}(y,x)\nonumber\\
-{\frac{1}{{\sqrt{r\bar r}}(1-{\frac{1}{r}})(px+qy)}}\bigg( u+v+(qx+py) {\frac{{V^\prime}_{n-2}}{V_{n-2}}} \bigg){V_{n+1}} \bigg]   \times \bigg[ \bigg( {\alpha_{xy}}+{\beta_{xy}} {\frac{{V^\prime}_{n-3}}{V_{n-3}}} \bigg) {({\omega_n})_{11}}(x,y)\nonumber\\
+ \bigg( {{\alpha^\prime}_{xy}}+{{\beta^\prime}_{xy}}{\frac{{V^\prime}_{n-3}}{V_{n-3}}} \bigg) {({\omega_n})_{22}}(x,y)- {\frac{1}{{\sqrt{r\bar r}}(1-{\frac{1}{r}})(qx+py)}}\bigg( u+v+(px+qy){\frac{{V^\prime}_{n-3}}{V_{n-3}}} \bigg) {V_n} \bigg]\nonumber\\
-{\frac{{D_{n-2}}{V_{n-4}}}{{Z^o_{n-2}}{Z^c_{n-4}}-{Z^c_{n-2}}{Z^o_{n-4}}}} \bigg[ \bigg( {\alpha_{yx}}+{\beta_{yx}} {\frac{{V^\prime}_{n-4}}{V_{n-4}}} \bigg) {({\omega_{n-1}})_{11}}(y,x)
+\bigg( {{\alpha^\prime}_{yx}}+{{\beta^\prime}_{yx}} {\frac{{V^\prime}_{n-4}}{V_{n-4}}} \bigg) {({\omega_{n-1}})_{22}}(y,x)\nonumber\\                            -{\frac{1}{{\sqrt{r\bar r}}(1-{\frac{1}{r}})(px+qy)}}\bigg( u+v+(qx+py) {\frac{{V^\prime}_{n-4}}{V_{n-4}}} \bigg) {V_{n-1}} \bigg] \hspace{0.5cm} 
\end{eqnarray}

The above equation reveals the recurrence relation among the partition functions of different Fibonacci generations from $(n-4)-th$ to $(n+2)-th$.The partition functions have entered in the above equation through the quantities given by equations from (48) to (57) .This is in conformity with the symmetry properties of the   Fibonacci chain.It is worth noticing that in the recurrence relation above ,the partition functions for both closed and open chains have entered.The procedure we have presented here through symmetry properties of the chain is generic to any substitution lattice. 

\vspace{0.2cm}

\noindent  {\bf  {IIIb Recurrence relation in the large $N$ limit}}

\vspace{0.1cm}

Recurrence relation (60) can be written in a much simpler form in the large $N$ limit.To achieve this we notice that in equation (60) both
open and closed partition functions of a particular generation occur simultaneously.From the physical consideration that thermodynamic
quantities must be same in the large $N$ limit for both open and closed         partition functions of a particular generation ,we assume that the free        
energy per spin is same in both cases in a particular generation.Therefore,      for the $n-th$ generation relation between the partition functions can be written as: 
\begin{equation}
{Z^c_n}(F)={f_n}(r,{\bar r},H) {Z^o_n}(F)
\end{equation}
where ${f_n}(r,{\bar r},H)$ is independent of the number of spins, ${N_n}+1$ ; $N_n$ being the number of bonds in the $n-th$ generation.

\vspace{0.2cm}

\noindent {\bf Case I:} {\bf Ising~ Limit} $\bf r=\bar r$ , $\bf H\ne 0$

\vspace{0.1cm} 

Considering equations (25) and (28) we can write equation (61) in the following form:

\begin{equation}
 {Z^c_{2M}}(F) {\vert_{\epsilon = {\bar \epsilon}}}\nonumber\\
={{[{\sqrt r}+{\sqrt r}(1-r){\frac{1}{{\lambda_+}^2}}+4(1-r){r^{\frac{-1}{2}}}{{cosh}^2}{\beta H}\times {\frac{1}{{{\lambda_+}^2}({{\lambda_+}^2}-{{\lambda_-}^2})}}]}^{-1}} {Z^o_{2M}}(F){\vert_{\epsilon ={\bar \epsilon}}}
\end{equation}
which is equivalent to equation (15) in the large $N$ limit where $2M$ stands 
for the number of spins in a large odd generation. From (61) and (62) we conclude that ${f^{odd}_n}(r={\bar r},H)$ is same for any large odd generation . Similarly, for large odd number of spins $2N+1$ in a large even generation, equations (26) and (29) give:
\begin{equation}
{Z^c_{2N+1}}(F){\vert_{\epsilon ={\bar \epsilon}}}={{[{\sqrt r}+4(1-r)cosh~{\beta H}\times {\frac{1}{{\lambda_+}({{\lambda_+}^2}-{{\lambda_-}^2})}}]}^{-1}}{Z^o_{2N+1}}(F){\vert_{\epsilon ={\bar \epsilon}}}
\end{equation}
which is equivalent to equation (16) in the large $N$ limit. From (61) and (63) we conclude that ${f^{even}_n}({r=\bar r},H)$ is same for any large even generation.

\vspace{0.2cm}

\noindent {\bf Case II:} {${\bf {r\ne {\bar r}},H=0}$}

\vspace{0.1cm}

Equations (25) and (28) give the exact expressions for open and closed partition
 functions with odd number of bonds in the following form:

\begin{equation}
{Z^o_{2M}}(F){\vert_{H=0}}={2^{2M}}{{(cosh~{\beta \epsilon})}^{N_L}}{{(cosh~{\beta {\bar \epsilon}})}^{N_S}}
\end{equation}
and
\begin{equation}
{Z^c_{2M}}(F){\vert_{H=0}}={2^{2M}}[{{(cosh~{\beta \epsilon})}^{{N_L}+1}}{{(cosh~{\beta {\bar \epsilon}})}^{N_S}}+{{(sinh~{\beta \epsilon})}^{{N_L}+1}}{{(sinh~{\beta {\bar \epsilon}})}^{N_S}}]
\end{equation}
It follows from equations (64) and (65) that in the large $N$ limit
\begin{equation}
{Z^c_{2M}}(F){\vert_{H=0}}=(cosh~{\beta \epsilon}) {Z^o_{2M}}(F){\vert_{H=0}}
\end{equation}
Equations (26),(29) give the exact expressions for open and closed partition functions with even number of bonds as:
\begin{equation}
{Z^o_{2N+1}}(F){\vert_{H=0}}={2^{2N+1}}{{(cosh~{\beta \epsilon})}^{N_L}}{{(cosh~
{\beta {\bar \epsilon}})}^{N_S}}
\end{equation}
and
\begin{equation}
{Z^c_{2N+1}}(F){\vert_{H=0}}={2^{2N+1}}[{{(cosh~{\beta \epsilon})}^{{N_L}+1}}{{(cosh~{\beta {\bar \epsilon}})}^{N_S}}-{{(sinh~{\beta \epsilon})}^{{N_L}+1}}{{(sinh~{\beta {\bar \epsilon}})}^{N_S}}]
\end{equation}
It follows from equations (67) and (68) that in the large $N$ limit
\begin{equation}
{Z^c_{2N+1}}(F){\vert_{H=0}}=(cosh~{\beta \epsilon}){Z^o_{2N+1}}(F){\vert_{H=0}}\end{equation}
From equations (61) , (66) and (69) we conclude that ${f^{odd}_n}({r\ne {\bar r}},H=0)$ [${f^{even}_n}({r\ne {\bar r}},H=0)$] is same for any large odd [even]
 generation . \\
Considering the above cases we make the assumption that in the case of Fibonacci with $H\ne 0$, all ${f_n^{even}}(r,{\bar r},H)$ in equation (61) for large odd generation number are equal and all ${f_n^{odd}}(r,{\bar r},H)$ in equation (61) for large even generation number are equal; but ${f_n^{even}}(r,{\bar r},H)\ne {f_n^{odd}}(r,{\bar r},H)$.\\ 
Using the above properties it reveals that the expression ${Z^o_{n-1}}{Z^c_{n-3}}-{Z^c_{n-1}}{Z^o_{n-3}}$ appearing in equation (60) vanishes.As a consequence equation (60) becomes:

\begin{eqnarray}
{{({\omega_{n+1}})}_{11}}(y,x)({\alpha_{yx}}+{\beta_{yx}}{\frac {V^{\prime}_{n-2}}{V_{n-2}}})
+{{({\omega_{n+1}})}_{22}}(y,x)({\alpha^{\prime}_{yx}}+{\beta^{\prime}_{yx}}{\frac {V^{\prime}_{n-2}}{V_{n-2}}}) \nonumber\\
-{\frac {1}{{\sqrt {r{\bar r}}}{(1-{\frac {1}{r}})}(px+qy)}}
\times {(u+v+(qx+py){\frac {V^{\prime}_{n-2}}{V_{n-2}}})}{({\frac {f^{even}_{n+1}}{\sqrt r}}-1)}{Z^o_{n+1}}=0
\end{eqnarray}

where

\begin{eqnarray}
{{({\omega_{n+1}})}_{11}}(y,x)
= [{\frac {{\Gamma_n}-{\Gamma^{\prime}_n}}{{\Gamma_n}{\Lambda^{\prime}_n}-{\Gamma^{\prime}_n}{\Lambda_n}}}\times {\frac {D_{n-1}}{r{\sqrt r}}}\times {\frac {\Delta_n}{Z^o_{n-2}}}](y,x)\times {Z^o_{n-2}}\nonumber\\
-{\frac {{\Gamma_n}{K^{\prime}}-{\Gamma^{\prime}_n}K{\frac {f^{even}_{n+1}}{\sqrt r}}}{
{\sqrt {r{\bar r}}}({\Gamma_n}{\Lambda^{\prime}_n}-{\Gamma^{\prime}_n}{\Lambda_n})}}(y,x)\times {Z^o_{n+1}}\nonumber\\
 = {{\Phi}^{11}_{n-2}}(y,x){Z^o_{n-2}}-{{\Phi}^{11}_{n+1}}(y,x){Z^o_{n+1}}
\end{eqnarray}
Here $\Phi (r,{\bar r},H)$'s are indepedent of the partition functions.Similarly , one can write
\begin{eqnarray}
{{({\omega_{n+1}})}_{22}}(y,x)={\Phi^{22}_{n-2}}(y,x){Z^o_{n-2}}-{\Phi^{22}_{n+1}}(y,x){Z^o_{n+1}}
\end{eqnarray}
In reducing ${{({\omega_{n+1}})}_{11}}(y,x)$ and ${{({\omega_{n+1}})}_{22}}(y,x)$ in the form (71) and (72) we have used the simplified expression
\begin{equation}
{\frac {V^{\prime}_n}{V_n}}={\frac {{\frac {f^{odd}_n}{\sqrt r}}-{\frac {1}{r}}}{{\frac {f^{odd}_n}{\sqrt r}}-1}}
\end{equation}
Since we have assumed $(n-2)-th$ generation to be even ;therefore $(n-1)-th$ , $n-th$ generations will be odd and $(n+1)-th$ generation will be even.\\ 
From equations (70),(71) and (72) we can write

$\{ {\Phi^{11}_{n+1}}(y.x)({\alpha_{yx}}+{\beta_{yx}}{\frac {V^{\prime}_{n-2}}{V_{n-2}}})+{\Phi^{22}_{n+1}}(y,x)({\alpha^{\prime}_{yx}}+{\beta^{\prime}_{yx}}{\frac {V^{\prime}_{n-2}}{V_{n-2}}})$\\

$+{\frac {1}{{\sqrt {r{\bar r}}}(1-{\frac {1}{r}})(px+qy)}}[u+v+(qx+py){\frac {V^{\prime}_{n-2}}{V_{n-2}}}]({\frac {f^{even}_{n+1}}{\sqrt r}}-1)\} {Z^o_{n+1}}$ \\     

$=\{ {\Phi^{11}_{n-2}}(y,x)+{\Phi^{22}_{n-2}}(y,x)\} {Z^o_{n-2}}$\\

or,
\begin{equation}
{\Psi_{n+1}}(r,{\bar r},H){Z^o_{n+1}}={\Psi_{n-2}}(r,{\bar r},H){Z^o_{n-2}}
\end{equation}
In terms of free energy per spin equation (74) takes the form
\begin{equation}
{({N_{n+1}}+1)}{F_{n+1}}-{({N_{n-2}}+1)}{F_{n+2}}={\beta}~log{\frac {\Psi_{n-2}}{\Psi_{n+1}}}
\end{equation}
Using the Fibonacci recurrence relation among the number of bonds expression (75) reduces to
\begin{equation}
(2\tau +1){F_{n+1}}={F_{n-2}}
\end{equation}
where ${\tau}={{lim}_{{N_n}\rightarrow \infty}}{\frac {N_L}{N_S}}$ , the golden mean . \\
Equation (76) shows that the free energy per spin for consecutive even generations are scaled by a factor $(2\tau +1)$ and therefore the thermodynamic quantities will be scaled in a similar fashion.

\vspace{0.2cm}

\noindent {\bf IIIc  Chemical Potential for H=0}

\vspace{0.1cm}

For calculating chemical potential we must know the exact form of the partition function . Since we do not know the exact partition function for finite $H$ we calculate it for $H=0$ .\\
Let $N_{n-2}$ denote the number of bonds in the $(n-2)-th$ generation which is even . We add another spin at the end with a long bond , say , to the open chain. Then using equation (26) we get
\begin{eqnarray}
{Z^o_{N_{n-2}+2}}(F){\vert_{H=0}}                                             
={r^{\frac {{N_L}+1}{2}}}{{\bar r}^{\frac {N_S}{2}}}Tr{{(1+{\frac {\sigma_1}{r}})}^{{N_L}+1}}{{(1+{\frac {\sigma_1}{\bar r}})}^{N_S}}({1+{\sigma_1}})\nonumber\\
={2^{{N_{n-2}}+2}}{{(cosh~{\beta \epsilon})}^{{N_L}+1}}{{(cosh~{\beta {\bar \epsilon}})}^{N_S}}
\end{eqnarray}
while the exact expression for the partition function with the open boundary condition in the $(n-2)-th$ generation is given in equation (67) as ,
\begin{equation}
{Z^o_{{N_{n-2}}+1}}{\vert_{H=0}}\nonumber\\
={2^{{N_{n-2}}+1}}{{(cosh~{\beta \epsilon})}^{N_L}}{{(cosh~{\beta {\bar \epsilon}})}^{N_S}}
\end{equation}
For large generation the expression for chemical potential as defined in equation (17) is 
\begin{equation}
{{e^{{\mu^o_{even}}(F)}}{\vert_{H=0}}}={\nu^o_{even}}(F){\vert_{H=0}}={\frac {{Z^o_{{N_{n-2}}+2}}(F)}{{Z^o_{{N_{n-2}}+1}}(F)}}=2cosh~{\beta \epsilon} (=2cosh~{\beta {\bar \epsilon}})
\end{equation}
The expression in the parenthesis stand for addition of a spin at the end of the chain with a short bond . \\
In a similar fashion , the chemical potential for closed chain is 
\begin{equation}
{\nu^c_{even}}=2cosh~{\beta \epsilon} (or 2cosh~{\beta {\bar \epsilon}})
\end{equation}
We would have got the same result had we considered odd generation. Comparing expressions (79) and (80) with that of (22) we conclude that , in absence of external magnetic field the chemical potential do not depend upon the underlying lattice structure as long as we consider nearest neighbour interaction. 

\vspace{0.5cm}

\noindent  {\bf {IV Conclusion}}

\vspace{0.2cm}

   We have found an exact expression for the partition function of an Ising 
model on a reguler lattice with open boundary conditions in presence of magnetic field .This always includes closed partition functions because of the fact that
 out of four different spin configurations at the end points of the chain ,  two configurations $\uparrow \uparrow$ and $\downarrow \downarrow$ satisfy closed boundary conditions.Free energy per spin are equal for both open and closed chains in the large $N$ limit , hence thermodynamic quantities are same in both cases ; but chemical potentials are different .For aperiodic chain we have established the recurrence relation among partition functions of different Fibonacci generations and we have also shown that this can only be obtained by using trace map relations and the characteristic symmetry property viz.,"Mirror Symmetry" of Fibonacci generations . This procedure is generic to all substitution lattices. In our case the recurrence relation among the partition functions correctly reflects the fact that $n-th$ and $(n\pm 6)-th$ generations are topologically same . As a consequence one must go through six times decimation renormalization group procedure to find scaling forms!
 of thermodynamic functions. The recurrence relation gets simplified in the large $N$ limit showing that free enrgy per spin in two consecutive large even generations are related through a scaling 
factor $2\tau +1$ . The expression for chemical potential in absence of magnetic field reveals that it is independent of the underlying lattice structure with nearest neighbour interaction .   

\vspace{0.5cm}

  Acknowledgement: We are thankful to S.N.Karmakar of SINP,Calcutta and to S.M. Bhattacharjee of IOP,Bhubaneswar for illuminating  discussions.One ofthe authors
(SKP) thanks J.P.Chakraborty of IACS,Calcutta for lending him necessary books from his library.

\vspace{1.0cm}

[1] Thomson Colin J 1972  $Phase$ $Transition$ $and$ $Critical$ $Phenomena$      

\vspace{0.1cm}

${\bf{Vol~I}}$ ed  by Domb C and Green M S (Academic Press , London)           

\vspace{0.1cm}

pp 177-225

\vspace{0.2cm}

[2] Stanley Eugene H 1971 $Phase$ $Transition$ $and$ $Critical$ $Phenomena$     

\vspace{0.1cm}

(Oxford University Press)

\vspace{0.2cm}

[3] Domb C 1974 Ising model $Phase$ $Transition$ $and$ $Critical$ $Phenomena$

\vspace{0.1cm}

${\bf{Vol~III}}$ ed Domb C and Green M S ( Academic Press , London )          

\vspace{0.1cm}

pp 357-458

\vspace{0.2cm}

[4] Grimm U and Baake M 1977 Aperiodic Ising Models $The$ $Mathematics$ $of$    

\vspace{0.1cm}

$Long$ $range$ $Aperiodic$ $Order$ $ed$ R M Moody (Doedrecht : Kluwer) 

\vspace{0.1cm}

pp 197-237

\vspace{0.2cm}

[5] Schechtman D , Blech I , Gratias D and Cahn J W 1984 $Phys$ $Rev$ $Lett$

\vspace{0.1cm}

$\bf 53$ 1951-1953

\vspace{0.2cm}

[6] Achian Yaakov , Lubensky T C and Marshall E W 1986 Ising model on a

\vspace{0.1cm}

quasiperiodic chain $Phys$ $Rev$ $\bf B33$ 6460-6464

\vspace{0.2cm}

[7] Tsunetsugu Hirokaju and Ueda Kazuo 1987 Ising spin system on the Fibonacci

\vspace{0.1cm}

chain $Phys$ $Rev$ $\bf B36$ 5493-5499
 
\vspace{0.2cm}

[8] Kohmoto Mahito , Kadanoff Leo P and Tang Chao 1983 Localization problem in  

\vspace{0.1cm}

one dimension : mapping and escape $Phys$ $Rev$ $Lett$ $\bf 50$ 1870-2

\vspace{0.2cm}

[9] Ostlund S , Pandit R , Rand D , Schellnhuber H J and Siggia E D 1983 One-

\vspace{0.1cm}

Dimensional Schrodinger Equation with an Almost Periodic Potential $Phys$ $Rev$ 

\vspace{0.1cm}

$Lett$ $\bf 50$ 1873-1876

\vspace{0.2cm}

[10] Baake M , Grimm U and Joseph D 1993 Trace Maps , invariants and some of 

\vspace{0.1cm}

their applications $Int$ $Jour$ $Mod$ $Phys$ $\bf B7$ 1527-1550

\vspace{0.2cm}

[11] Nang Xiaoguang , Grimm Uwe and Screiber Michael 2000 Trace and antitrace   

\vspace{0.1cm}

maps for aperiodic sequences , their extensions and applications 

\vspace{0.1cm}

$cond-mat$/$0005463$

\vspace{0.2cm}

[12] Huang Kerson 1987 $Statistical$ $Mechanics$ Second Edition                 

\vspace{0.1cm}

(John Wiley Sons)

\end{document}